# DeepThalamus: A novel deep learning method for automatic segmentation of brain thalamic nuclei from multimodal ultra-high resolution MRI

Marina Ruiz-Perez[1], Sergio Morell-Ortega[1], Marien Gadea[2], Roberto Vivo-Hernando[3], Gregorio Rubio[4], Fernando Aparici[5], Mariam de la Iglesia-Vaya[6], Thomas Tourdias[7], Boris Mansencal[8], Pierrick Coupé[8] and José V. Manjón[1]

[1] Institute of Applications of Information and Advanced Communication Technologies (ITACA), Polytechnic University of Valencia, Valencia, Spain.

[2]Department of Psychobiology, Faculty of Psychology, University de Valencia, Valencia, Spain

[3] Institute of Automation and Industrial Informatics, Polytechnic University of Valencia, Valencia, Spain.

[4] Department of Applied Mathematics, Polytechnic University of Valencia, Valencia, Spain.

[5] Medical Imaging Area. La Fe University and Polytechnic Hospital. Valencia, Spain.

[6] Mixed Unit of Biomedical Imaging FISABIO-CIPF. Foundation for the Promotion of Health and Biomedical Research of the Valencian Community - Valencia, Spain.

[7] University Bordeaux, INSERM, Neurocentre Magendie, U1215, Bordeaux, France

[8] CNRS, University Bordeaux, Bordeaux INP, LABRI, UMR5800, in2brain, Talence, France

**Corresponding author:** E-mail: mruiper@etsii.upv.es

# Abstract

The implication of the thalamus in multiple neurological pathologies makes it a structure of interest for volumetric analysis. In the present work, we have designed and implemented a multimodal volumetric deep neural network for the segmentation of thalamic nuclei at ultra-high resolution (0.5 mm isotropic). Most current tools either operate at standard resolution (1 mm isotropic) or use monomodal data. To build the roposed method, first, a database of semiautomatically segmented thalamic nuclei was created using ultra-high resolution T1, T2 and White Matter nulled (WMn) images. Then, a novel Deep learning-based strategy was designed to obtain the automatic segmentations trained using a novel semi-supervised approch to improve its robustness and accuracy. The proposed method was compared with a related state-of-the-art method demonstrating competitive results both in terms of segmentation quality and efficiency. To ensure our method's accessibility to the scientific community, a full pipeline able to work with monomodal standard resolution T1 images is also proposed.

# 1. Introduction

The thalamus plays an important role in various neurological disorders. In Parkinson's disease (PD), the volumes of the thalamus and the thalamic subregions are lower even in drug-naïve patients [1]. In other dementias like Alzheimer's disease (AD) [2] or Lewy body dementia [3] thalamic atrophy has also been observed. Moreover, demyelinating diseases like progressive Multiple Sclerosis (MS) [4] or psychiatric ones like chronic Schizophrenia [5] also exhibit some degree of thalamic atrophy. Therefore, comprehending the anatomy of the thalamus holds significant potential for characterizing the neurological state of the human brain.

The thalamus serves as a critical hub within the cortico-striatal-thalamo-cortical circuit (CSTC), a neural network crucial for regulating movement execution, habituation, and reward processing [6]. It is made of multiple nuclei projecting to distinct cerebral areas and participating in diverse activities. The thalamus nuclei are categorized into various groups based on their anatomical placement and connections [7]. Given the importance of the thalamus in numerous pathological conditions, it is of capital interest to quantitatively assess its volumetric patterns through the development of automated segmentation algorithms that accurately define thalamic nuclei in vivo, enabling the identification of reliable and precise biomarkers.

Several segmentation methods have been proposed for this task based on structural magnetic resonance images (MRI). For whole thalamus segmentation, multi-atlas-based methods have been proposed [14,15,16] and, more recently, deep learning-based ones [17,18]. There are several available image processing software packages that segment the thalamus from MRI data such as FreeSurfer [13], FSL-FIRST [14] or volBrain [9]. However, to have a more detailed insight into the thalamus sub-structure, new methods for thalamic nuclei segmentation have been proposed. For example, there are segmentation methods that register histologically derived labels to MRI [15]. However, these methods are known to be very sensitive to the quality of the registration process. As commented, FreeSurfer also offers thalamic nuclei segmentation capabilities [16] based on a probabilistic atlas combining ex-vivo MRI and histology. There are also methods for thalamic nuclei segmentation based on functional data such as diffusion MRI or functional MRI (fMRI). For example, using diffusion MRI data, it has been proposed an interesting solution based on the improved contrast of this MRI modality (at the expense of a lower resolution) [17]. More recently, innovative works such as [18] describe a new manual delineation protocol based on optimized super-resolution (0.25 mm) short-tracks TDI, an imaging method based on diffusion tractography. The use of fMRI also provides connectivity information for the segmentation of thalamic nuclei [25,26], using resting state fMRI, used to examine functional relations of cortical areas using Independent Component Analysis. Unfortunately, these fMRI-based methods have an inherent low resolution which limits their usefulness.

Recently, the development of new MRI acquisition sequences has improved segmentation quality using structural MRI. White-Matter-nulled (WMn) images have significantly boosted the thalamic image contrast, allowing for a more precise segmentation. Methods such as THOMAS [21] utilize multi-atlas-based approaches, while other methods, like [22], apply deep learning techniques. In [29] a 2.5D slice-wise patch-based sliding window approach in native image space is used to segment an MPRAGE volume (they also presented a deep learning-based method using synthesized WMn images). Using also image synthesis, [23] proposes HIPS, a transformation step (histogram-based polynomial synthesis) into the THOMAS pipeline for WMn image synthesis, using a polynomial approximation for intensity transformation.

Taking into consideration all the mentioned existing methods, we can conclude that most of them work at standard resolution (at most 1 mm isotropic) and work on monomodal data (usually T1 or WMn). Given the size of some thalamic nuclei, it is crucial to use higher resolution data and given the low contrast of commonly used T1 MRI images it seems also necessary to use multimodal data to avoid thalamus volume overestimation through a richer feature characterization.

In this paper, we propose a novel method using deep learning for thalamic nuclei segmentation that works with multimodal data at ultra-high resolution (0.5 mm isotropic). In the following sections, the details of the proposed method are presented including a novel incremental semi-supervised deep learning-based learning method. Finally, a fully automatic segmentation pipeline for thalamic nuclei segmentation able to work with standard resolution T1 images is also presented.

# 2. Material and methods

The proposed method in this paper has been created using two datasets at different stages of its development.

- **HCP dataset**: The first one is a subset of MR images from the Human Connectome Project (HCP) [24]. HCP dataset consists of MR images taken on a 3T machine from 1200 healthy subjects aged between 22 and 35 years. Specifically, it includes high resolution T1 and T2 images (matrix size = 260 x 311 x 260 voxels, voxel size = 0.7 mm isotropic). We used a subset of 75 subjects containing T1, T2 and Synthetic WMn images and the thalamic nuclei segmentation labels. Details of this dataset can be found in [25].

- **Lifespan dataset**: The second dataset it is made of set of quality curated 4856 standard resolution T1 images from various public databases previously used in a previous project [26] to construct a lifespan model of the human brain with men and women from 3 to 90 years old, 2887 were healthy controls and 1969 were pathological (AD, ASD, and others). The details of these databases are as follows:

  1. **C-MIND (N=236):** All the images were acquired at the same site on a 3T scanner. The MRI data comprised 3D T1 MPRAGE high-resolution anatomical scan of the entire brain with a spatial resolution of 1 mm isotropic (https://research.cchmc.org/c-mind/).
  2. **NDAR (N=493):** The National Database for Autism Research (NDAR) is a national database funded by NIH (https://ndar.nih.gov). This database includes 13 different MRI cohorts acquired on 1.5T and 3T scanners.
  3. **ABIDE (N=905):** The images from the Autism Brain Imaging Data Exchange (ABIDE) dataset (http://fcon_1000.projects.nitrc.org/indi/abide/) were obtained on 905 subjects acquired at 20 different sites on 3T scanners.
  4. **ICBM (N=294)**: The images from the International Consortium for Brain Mapping (ICBM) dataset (http://www.loni.usc.edu/ICBM/) were obtained on 294 subjects through the LONI website.
  5. **OASIS (N=393):** The 393 control subject images came from the Open Access Series of Imaging Studies (OASIS) database (http://www.oasis-brains.org).
  6. **IXI (N=549)**: The images from the Information eXtraction from Images (IXI) database (http://brain-development.org/ixi-dataset) consist of 549 normal subjects from 1.5T and 3T scanners.

7. **ADNI (N=1649):** Data used in the preparation of this article were obtained from the Alzheimer's Disease Neuroimaging Initiative (ADNI) database (adni.loni.usc.edu). The ADNI was launched in 2003 as a public-private partnership, led by Principal Investigator Michael W. Weiner, MD. The primary goal of ADNI has been to test whether serial magnetic resonance imaging (MRI), positron emission tomography (PET), other biological markers, and clinical and neuropsychological assessment can be combined to measure the progression of mild cognitive impairment (MCI) and early Alzheimer's disease (AD). For up-to-date information, see www.adni-info.org. This dataset includes 1649 subjects from the 1.5T baseline collection. These images were acquired on 1.5T MR scanners at 60 different sites across the United States and Canada.
8. **AIBL (N=337)**: The Australian Imaging, Biomarkers and Lifestyle (AIBL) database (http://www.aibl.csiro.au/) used in this study consists of 337 subjects. The imaging protocol was defined to follow ADNI's guideline on the 3T scanner and an MPRAGE sequence on the 1.5T scanner.

## 2.1. Preprocessing

The images of the HCP dataset passed through a preprocessing stage to place them into a standard geometric and intensity space. This stage prepares the data for processing and consists of the following steps:

- **Noise removal**: The Spatially Adaptive Non-local Means (SANLM) filter [27] was used to reduce random noise naturally present in the images.
- **Inhomogeneity correction**: The N4 bias correction method was applied to correct the inhomogeneity of the images due to the acquisition process [28].
- **Registration**: Affine registration was performed to align the images with the ultra-high resolution MNI152 space (0.5mm isotropic resolution). ANTs software [29] was employed for this task. The resulting images have a standard matrix size of 362x434x362 voxels and a resolution of 0.5 mm isotropic.
- **Area of interest cropping**: Once the images were placed in MNI152 space, they were cropped to select only the sub volume containing the right and left thalamus. Cropped volumes were obtained using predefined limits based on the thalamus labels with a security margin of 10 voxels in each dimension to account for anatomic position variability. The final cropped volume had a size of 76x91x79 voxels.

In summary, the final library, contains 150 (75 left and 75 right thalamic regions) T1, T2 and synthetic WMn images with their corresponding labels (right crops were flip to obtain a total of 150 left crops). This dataset was divided into three subsets, training (N=120), validation (N=10) and test (N=20). Left and right regions of the same case were included in the same subset to avoid data contamination.

## 2.2. Neural Network Architecture

In deep learning, the selection of the proper architecture is of key importance to ease the learning process. In medical image segmentation, probably the most well-known and used architecture is the U-Net model proposed in 2015 at the MICCAI conference [30] for the segmentation of biomedical images. This model is a Fully Convolutional Network (FCN) comprising an encoder and a decoder with skip connections. The encoder captures image context and comprises a series of convolutional and pooling layers, effectively reducing the image dimensions. In this work, we evaluated a 3D version. The used 3D U-Net model comprises 4 resolution blocks, each one with two sub-blocks consisting of a 3D Convolutional layer followed by a ReLU activation and a batch normalization layer. Prior to down-sampling, a dropout layer is applied in each resolution block. Down-sampling was performed using strided convolution (stride=2). In the decoder, linear resizing layers were used to merge features from higher resolution branches with lower resolution ones through concatenation, facilitating information flow across resolutions. No dropout was used in the decoder. The final output is obtained through a SoftMax activation function applied to the last 3D convolutional layer. The number of filters varies dynamically throughout the network. In the encoder, the filter count starts at 56 and doubles with each subsequent layer (56,112, 224, 448), while in the decoder, it begins at 224 and halves at each subsequent layer (all convolutions used a 3x3x3 voxels kernel size). We used this U-Net definition as we used it in previous projects with satisfactory results. However, we are aware that other U-Net configurations are possible (i.e. nnU-Net [31]).

Although the U-Net model is a popular choice, its complexity can increase the risk of overfitting and reduce generalizability to unseen data. Based on the principle of parsimony, simpler models with fewer parameters are often preferred if they perform similarly, as they may lower overfitting risks while maintaining or even enhancing generalizability. Consequently, exploring alternative, simplified architectures could provide an effective and efficient balance between model capacity and the potential for overfitting.

One potential alternative architecture is to exclusively employ a decoder-based network, bypassing the encoder component. In pursuit of this goal, we propose the Deep Pyramidal Network (DPN) which shares some similarities to other pyramid based architectures [32]. Specifically, our proposed architecture employs an incremental approach, consisting solely on the decoder part, resulting in fewer parameters and reduced computational overhead. The network starts with an input tensor, representing the original high-resolution image (0.125 mm isotropic in our case). By employing Average Pooling layers, inputs at resolutions of 1/2, 1/4, and 1/8 of resolution are consecutively derived. The network starts from a 1/8 resolution input, obtained by subsampling the original volume by a factor of 8. Three convolution blocks with ReLU and Batch Normalization are applied to this input with a final Droput layer at the end of the block. The resulting tensor is then upsampled using linear resize layer to match the dimensions of the 1/4 resolution convolved input, aiming to extract high-frequency features, which are subsequently concatenated. This process is repeated until the original resolution is reached (with the exception of the dropout layer at the original resolution level). A Convolutional Block Attention Module (CBAM) [33] block is used at the end of the 1/4 resolution level. The final tensor undergoes a processing through a softmax activation layer, yielding the model's output tensor. Each convolution employs a 3x3x3 kernel. The DPN network maintains a consistent number of filters across all resolution blocks, specifically, 56 in this project. The fundamental concept underlying this architecture is its incremental approach, where the initial layers generate broad patterns that are progressively refined as the network scales in resolution. Figure 1 shows a multimodal version of the proposed architecture.

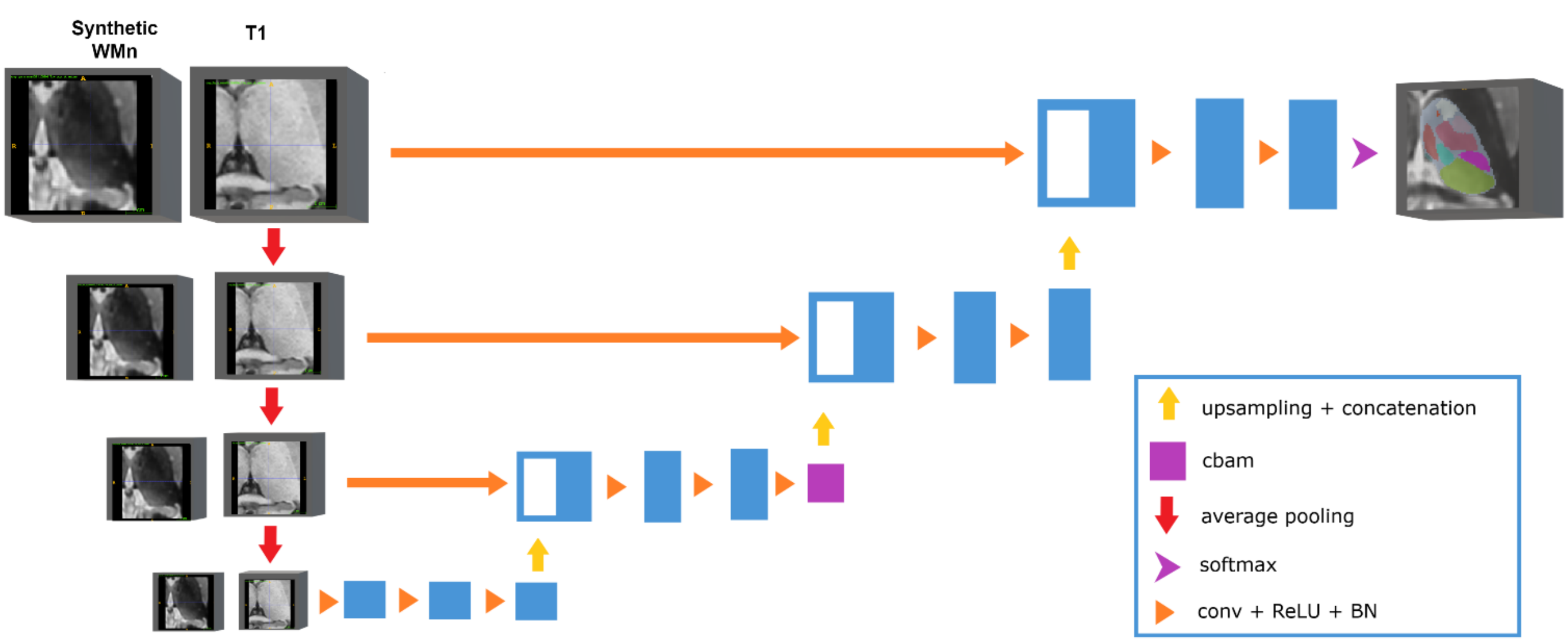

*Figure 1. Proposed multimodal DPN segmentation architecture applied to a multimodal input (T1 and WMn).*

## 2.3. Atlas prior creation

In addition of using multimodal input data, we also explored the utilization of *a priori* information into the input of the neural network in form of an atlas as done in previous approaches [16] to enhance the model robustness. This atlas can introduce a priori information of the expected location and size of the thalamic nuclei which may make the method more robust to size and location variations during the whole human lifespan and combine the benefits of atlas based and deep learning methods.

We propose to construct a subject-specific atlas using an ultra-fast deep learning based non-linear registration method combined with a multi-atlas label fusion approach. Specifically, we trained a convolutional neural network (CNN) for unsupervised non-linear registration using an approach similar the popular Voxelmorph network [34]. We trained it using the cropped T1 images of the library. After training the network, it was used to register the 20 most similar cases to the case to be segmented (excluding it) to create the subject-specific atlas using the mean absolute error metric. A voxel wise local weighted majority voting algorithm is employed for label fusion, prioritizing labels from the library with intensities that closely resemble the image being segmented. We did not use a non-local label fusion [35] which is much more powerful to limit the computational load of the method (the whole atlas creation process takes around 15 seconds). This atlas is integrated as an additional input channel into the architecture, alongside the input MR images.

## 2.4. Training process

Experiments with different architectures (U-Net and DPN) and with different settings of input data were performed and the best results were selected. To train the different networks we used a loss function combining a variant of the generalized dice loss (GDL) [36] and binary cross entropy. Differently from the GDL, where they weighed each label by the inverse of the volume of each case label, we use the inverse of the mean volume of the label in the training dataset (this way at each step the loss function is more stable). Binary cross entropy was added to the GDL as it has shown in previous works that this combination improves the results. Combining dice-based loss with entropy related losses has become lately a standard in medical image segmentation [39].

We used the logarithm of the loss function rather than the loss itself to increase the magnitude of the gradients at convergence and prevent early stopping. The used loss function is shown in Equation 1.

$$\mathrm{loss} = \log\left(\mathrm{GDL}(\mathrm{y}, \mathrm{p}) + BCE(y, p) + \mathrm{eps}\right) \tag{1}$$

where *p* is the predicted probability, *y* the true probability and eps is a small value to avoid zero logarithm values. All models were trained for 1500 epochs with 50 steps per epoch. The Adamax optimizer, a variant of the Adam optimizer [37] has been used to train the networks, instead of using the second order moving average of the gradients, it uses the absolute maximum value of the cumulative gradients to normalize the weight updates. Data augmentation was used to improve method generality. This was done through geometric and intensity transformations, facilitated by TorchIO [38]. Geometric transformations such as resizing, rotations, and deformations enable the model to comprehend varying spatial arrangements. Simultaneously, intensity transformations, involving contrast adjustments, variations in brightness, and the introduction of controlled noise were used. Both networks have dropout layers (rate=0.2) at the first layers which helps prevent overfitting. Input images and the atlas were normalized using z-scoring.

To evaluate the segmentations, the Dice index was used, which quantifies the similarity between the actual segmentation and the segmentation performed by the algorithm. The mean DICE values of the 13 nuclei of the thalamus, the DICE value for each nucleus and the DICE value of the whole thalamus were obtained. We also used the percent Volume Difference (VD) described in Equation 2.

$$\text{VD} = 100(Vg - Vp)/Vg \tag{2}$$

where *Vg* is the true volume and *Vp* the predicted volume.

## 2.5. Semi-supervised learning

It is fundamental for any segmentation model to be applicable to cases with different ages, pathologies, or any other image conditions. In other words, the method must be robust to process out-of-domain cases without a noticeable loss in accuracy. We are aware that the proposed model trained using only the DeepThalamus dataset (with images from subjects aged between 22 and 36 years) will likely poorly perform in subjects outside of the training age range (the model may potentially suffer from lack of generalization errors). For this reason, we designed a procedure to adapt the method to age-and pathology-related variations in thalamic shape using an advanced preprocessing and the use of a subject-specific atlas. However, this approach may be insufficient in certain scenarios.

Therefore, we decided to expand the training dataset by conducting fine-tuning using a semi-supervised approach, utilizing the lifespan dataset previously described. This dataset is composed of 4856 unlabeled cases covering the whole lifespan and several pathologic anatomical patterns. The simplest approach would be to apply the trained network to the whole lifespan dataset and use the resulting segmentations as pseudo-labels to further train the proposed method. However, this simple approach is likely to introduce errors and degrade the method´s performance severely (cases like those used in training are probably producing good quality segmentations but very different cases will likely result in poor quality segmentations).

To avoid this problem, we propose to use an incremental approach like POPCORN [39] used for atlas propagation in the context of a multi-atlas segmentation. Specifically, we propose to first select the *N* most similar cases of the lifespan dataset to the training dataset and segment them with the current network (expecting a good quality output due to their high similarity to the training dataset) and later add them (with their automatic segmentations) to the training dataset to train the network with this extended dataset. We repeat this process using *N* size batches until the whole lifespan dataset is included in the training dataset. At each iteration, the last fine-tuned version of the network is used to segment the new batch of *N* subjects. This approach reduces the *out-of-domain* gap and ensures a smooth progression through the extended dataset.

To estimate the similarity between cases, we used their projections into a lower-dimensional space using the UMAP algorithm [40]. Specifically, we first trained a convolutional autoencoder using the DeepThalamus and lifespan datasets at MNI space (T1 images only). The autoencoder serves as a preliminary step for data compression and feature extraction, reducing the dimensionality of 3D images and preparing them for UMAP, which performs a non-linear projection of those latent features into a lower-dimensional space, easing the analysis of the images. Once the autoencoder was trained, the latent space of each case was estimated to obtain a compact representation in both datasets. Later, for visualizing and calculating of the proximity graph, the UMAP algorithm was used. This algorithm allows us to project the data in 2 dimensions and to calculate the distances in this space. One of the main advantages of this algorithm is that it is able to preserve the topology of multidimensional data embeddings, which allows to calculate the distances in the projected space. Figure 2 shows the UMAP projection, labeled according to age evolution, source dataset and diagnosis. In the visualization by age, a clear progression from younger to older ages can be observed.

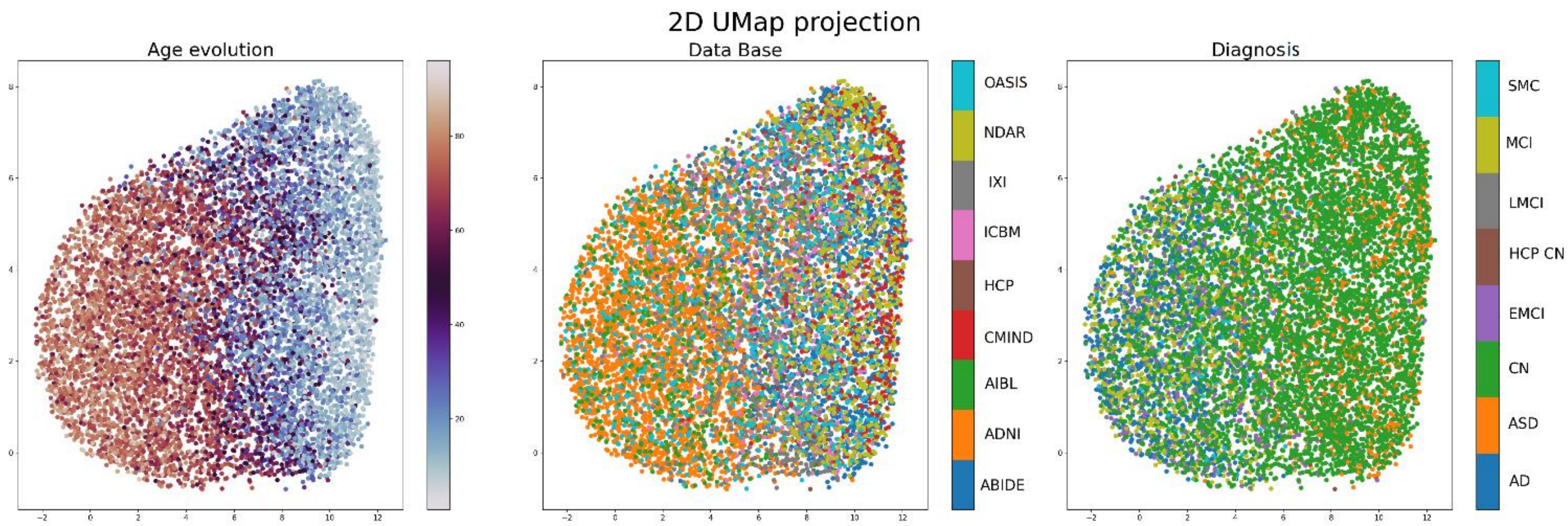

*Figure 2. 2-D projection of the lifespan dataset as a function of age, database and diagnosis with the UMAP algorithm. The left graph shows the projection of the data as a function of age, where the youngest and oldest people are situated around middle-aged. The center one is a function of the database, showing that the different databases form clusters. The right graph shows the data as a function of the diagnosis, where the subjects with AD are far from the control and in between there are other pathologies.*

To select the *N* most similar cases at each batch a K-nearest neighbor algorithm (k = 50) was used. The lifespan dataset comprises 4856 images, encompassing 9712 thalami (left and right). The model was configured for 50 iterations, incorporating N=197 new subjects at each iteration. To avoid balancing problems (lifespan is bigger than DeepThalamus dataset) the contribution of each dataset is balanced. The proportions of each dataset used in the training were, in the first iteration, 50% for the HCP dataset and 50% for the first 197 lifespan images, from the second iteration to the last, the probability was 50% for the HCP dataset, 25% for the 197 new lifespan images and 25% for the previously incorporated lifespan subjects at previous iterations.

To use the proposed method to segment the lifespan dataset we had to increase the resolution of the T1 images from 1 mm isotropic to 0.5 mm isotropic. To do so, we used an in-house super-resolution algorithm based on a 3D ResNet architecture and trained with the HCP dataset. To synthesize the WMn images from the T1 images we used the synthesis network used in [25] .

# 3. Results

In this section, we summarize the results of the experiments performed to obtain the best segmentation network based on the different architectures and input data options previously described. In the next sections, results of these experiments are shown to justify the decisions taken on the method design and implementation. All the experiments were performed with TensorFlow 1.15, using a NVIDIA V100 32GB GPU, an Intel Xeon processor and with 64 GB RAM; all running under Linux Ubuntu 18.04. All networks were trained for 1500 epochs (one day of processing) assuring that their convergence was reached. The resulting networks were evaluated using the test subset of the described HCP dataset (N=20) using the dice coefficient and VD metric.

## 3.1. Architecture selection

One of the first experiments we performed was the comparison of the two described architectures (U-Net and DPN). Both networks were trained and tested using the HCP dataset using the same multimodal input (T1 and WMn images). We present the results using T1 and WMn as the input images, as this configuration yielded the best outcomes, as indicated in Table 3. The result is shown in Table 1. Statistical differences were tested using the two-sided Wilcoxon signed-rank test ($p<0.05$). Because multiple comparisons were performed, a correction was applied to control the false discovery rate (FDR) using the Benjamini-Hochberg method. It is noteworthy that DPN architecture has 1.034.355 trainable parameters while U-Net has 8.394.470 (i.e. eight times less).

*Table 1. Average Dice results for the whole thalamus and the 13 thalamic nuclei for the U-Net and DPN networks. Best results in bold (*represents statistically significant differences).*

| | DPN | U-Net |
|---|---|---|
| Whole thalamus | **0.9795 ± 0.0041*** | 0.9751 ± 0.0040 |
| *Anterior Ventral Nucleus* | **0.9169 ± 0.0298** | 0.9008± 0.0340 |
| *Ventral Anterior Nucleus* | **0.9211± 0.0220** | 0.9125 ± 0.0287 |
| *Ventral Lateral Anterior Nucleus* | **0.9144 ± 0.2080** | 0.8978± 0.0402 |
| Ventral Lateral Posterior Nucleus | **0.9578 ± 0.0090*** | 0.9447 ± 0.0145 |
| Ventral Posterior Lateral Nucleus | **0.9403 ± 0.0151*** | 0.9034 ± 0.0265 |
| *Pulvinar Nucleus* | **0.9721 ± 0.0052*** | 0.9616 ± 0.0078 |
| *Lateral Geniculate Nucleus* | **0.9081 ± 0.0258*** | 0.8980 ± 0.0360 |
| *Medial Geniculate Nucleus* | **0.9385 ± 0.0145** | 0.9295 ± 0.0294 |
| *Centromedian Nucleus* | **0.9505 ± 0.0134*** | 0.9333 ± 0.0233 |
| *Mediodorsal Nucleus* | **0.9719 ± 0.0073*** | 0.9635 ± 0.0120 |
| *Habenular Nucleus* | **0.9082 ± 0.0374** | 0.9055 ± 0.0337 |
| *Mammillothalamic Tract* | **0.8705 ± 0.1076*** | 0.8531 ± 0.1139 |
| *Intermediate Space* | **0.9028 ± 0.0138*** | 0.8774 ± 0.0163 |
| Average | **0.9287 ± 0.0456** | 0.9142 ± 0.0538 |

The best results correspond to the DPN network, despite its lower complexity (9 of the 15 measures were statistically better for the DPN network). Given the small size of the training set (N=120), this network seems not to produce as much overfitting as the U-Net due to its smaller number of parameters. Based on these results we chose the DPN architecture for the development of our method.

## 3.2. Resolution analysis

Another hypothesis of our proposed approach is that working at higher resolution improves the method's accuracy. To evaluate this hypothesis, we trained the proposed DPN network at 0.5 mm and 1 mm isotropic resolutions (down-sampling by a factor 2 the original HCP dataset) using T1 and synthetic WMn images. The results are shown in Table 2. As can be noted, the best results were obtained at high resolution as expected.

*Table 2. Comparison of average Dice results for high and standard resolution images and the volume difference (VD). Best results in bold (*represents statistically significant differences).*

| | Dice | | VD (%) | | |
|---|---|---|---|---|---|
| | **High resolution (0.5 mm)** | **Standard resolution (1 mm)** | **High resolution (0.5 mm)** | **Standard resolution (1 mm)** | **Voxels (0.5 mm)** |
| Whole thalamus | **0.9795 ± 0.0041*** | 0.9768 ± 0.0034 | **0.8392 ± 0.8513** | 1.1136 ± 0.9100 | 32892 |
| *Anterior Ventral Nucleus* | **0.9169 ± 0.0298*** | 0.9031± 0.0404 | **5.1070 ± 5.0103** | 7.4142 ± 9.7199 | 574 |
| *Ventral Anterior Nucleus* | **0.9211 ± 0.0220*** | 0.9103 ± 0.0356 | **3.3970 ± 3.1514*** | 5.1700 ± 3.0870 | 1440 |
| *Ventral Lateral Anterior Nucleus* | **0.9144 ± 0.2080*** | 0.8963 ± 0.0321 | **7.0871 ± 4.4052** | 9.2832 ± 6.7694 | 493 |
| Ventral Lateral Posterior Nucleus | **0.9578 ± 0.0090*** | 0.9496 ± 0.0123 | **1.6526 ± 1.7383** | 2.8713 ± 2.6838 | 4518 |
| Ventral Posterior Lateral Nucleus | **0.9403 ± 0.0151*** | 0.9231 ± 0.0163 | **1.5124 ± 1.1494*** | 4.5922 ± 2.6128 | 1918 |
| *Pulvinar Nucleus* | **0.9721 ± 0.0052*** | 0.9662 ± 0.0070 | **1.0115 ± 0.8213** | 1.0757 ± 0.7635 | 7294 |
| *Lateral Geniculate Nucleus* | **0.9081 ± 0.0258*** | 0.9022 ± 0.0391 | **4.8612 ± 3.8705** | 6.6099 ± 4.1399 | 463 |
| *Medial Geniculate Nucleus* | **0.9385 ± 0.0145*** | 0.9316 ± 0.0252 | **4.2842 ± 2.6369** | 6.6696 ± 4.1063 | 427 |
| *Centromedian Nucleus* | **0.9505 ± 0.0134*** | 0.9295 ± 0.0253 | **3.5647 ± 2.7376*** | 6.0686 ± 3.1594 | 704 |
| *Mediodorsal Nucleus* | **0.9719 ± 0.0073*** | 0.9665 ± 0.0060 | **1.8156 ± 1.2522** | 2.3042 ± 1.6773 | 3188 |
| *Habenular Nucleus* | **0.9082 ± 0.0374*** | 0.8901 ± 0.0653 | **6.2531 ± 4.6660** | 9.5006 ± 0.2258 | 124 |
| *Mammillothalamic Tract* | **0.8705 ± 0.1076*** | 0.8105 ± 0.1540 | **9.5563 ± 12.399*** | 19.4326 ± 6.393 | 112 |
| *Intermediate Space* | **0.9028 ± 0.0138*** | 0.8840 ± 0.0163 | **2.4717 ± 2.3465*** | 3.8709 ± 2.6875 | 11636 |
| Average | **0.9287 ± 0.0456*** | 0.9123 ± 0.0647 | **4.0441 ± 5.1684*** | 6.5279 ± 9.9119 | |

For Dice coefficient, all the measures were statistically better for HR compared to LR. For the VD metric, all measures were lower for HR compared to LR being significantly different 6 of them. We provide the average size of each structure as reference as the dice coefficient is sensitive to the structure size[41]. This is because dice score is based on the ratio of overlap between the predicted and true labels, which is influenced by the number of pixels in the regions being compared. For VD metric, since most segmentation errors tend to be in the object's boundaries the reduced surface/volume ratio at HR favors the reduction of volume estimation errors.

## 3.3. Modality selection

We also explored what multimodality setting was the best performing one. We compared all possible combinations of inputs, both monomodal and multimodal with the proposed DPN network. Results are summarized in Table 3.

*Table 3. Average Dice results for the different modalities. Best result in bold.*

| Modalities | Dice |
|---|---|
| T1 | 0.8993 ± 0.0601 |
| T2 | 0.8548 ± 0.0872 |
| Synthetic WMn | 0.9264 ± 0.0418 |
| T1 + T2 | 0.9037 ± 0.0614 |
| T1 + Synthetic WMn | **0.9287 ± 0.0456** |
| T2 + Synthetic WMn | 0.9250 ± 0.0483 |
| T1 + Synthetic WMn + T2 | 0.9249 ± 0.0506 |

From the experiments performed with monomodal inputs, we can conclude that the best segmentation was obtained using the synthesized WMn images (0.9264). This result is consistent with recent literature [28,29,53], where the best modality for the task was found to be WMn. This high performance can be attributed to the fact that WMn images offer higher contrast between thalamic nuclei, which supports their clinical relevance. These results highlight the importance of obtaining WMn images directly or through advanced image synthesis models.

When combining the synthesized WMn images with T1 images, the overall Dice value improves to 0.9287, confirming our initial hypothesis that multimodal inputs provide more information than a single modality. Positive effects are also seen when combining the T1 modality with T2, where the joint segmentation reaches a higher value than when using these images individually (0.9037 vs. 0.8993 and 0.8548, respectively). This reinforces again the idea that combining modalities improves segmentation results compared to using single-modality images.

Finally, when adding all three modalities together (T1, T2 and synthesized WMn), the Dice value does not improve compared to T1+WMn, suggesting that the inclusion of multiple modalities does not always add useful additional information. In this case, the T2 combination may not be contributing significantly to thalamus segmentation but introducing noise in the learning process.

Comparing in detail the two best configurations (T1 + WMn synthesized and WMn synthesized alone), we can conclude that the combination of T1 + WMn synthesized offers superior performance in most of the segmented structures, being statistically significant in four of them (see Table 4).

*Table 4. Thalamus nuclei Dice results for the 2 best performing options. Best results in bold (*represents statistically significant differences).*

| | T1 + Synthetic WMn | Synthetic WMn |
|---|---|---|
| Whole thalamus | **0.9795 ± 0.0041*** | 0.9761 ± 0.0044 |
| *Anterior Ventral Nucleus* | 0.9169 ± 0.0298 | **0.9210 ± 0.0342** |
| *Ventral Anterior Nucleus* | 0.9211 ± 0.0220 | **0.9255 ± 0.0211** |
| *Ventral Lateral Anterior Nucleus* | **0.9144 ± 0.0208** | 0.9036 ± 0.0301 |
| Ventral Lateral Posterior Nucleus | **0.9578 ± 0.0090** | 0.9552 ± 0.0099 |
| Ventral Posterior Lateral Nucleus | **0.9403 ± 0.0151*** | 0.9272 ± 0.0206 |
| *Pulvinar Nucleus* | **0.9721 ± 0.0052*** | 0.9687 ± 0.0065 |
| *Lateral Geniculate Nucleus* | **0.9081 ± 0.0258** | 0.8971 ± 0.0363 |
| *Medial Geniculate Nucleus* | 0.9385 ± 0.0145 | **0.9399 ± 0.0133** |
| *Centromedian Nucleus* | **0.9505 ± 0.0134** | 0.9492 ± 0.0151 |
| *Mediodorsal Nucleus* | **0.9719 ± 0.0073** | 0.9710 ± 0.0076 |
| *Habenular Nucleus* | 0.9082 ± 0.0374 | **0.9099 ± 0.0380** |
| *Mammillothalamic Tract* | 0.8705 ± 0.1076 | **0.8782 ± 0.1018** |
| *Intermediate Space* | **0.9028 ± 0.0138*** | 0.8971 ± 0.0143 |
| Average | **0.9287 ± 0.0456** | 0.9264 ± 0.0418 |

## 3.4. Use of a priori information

Finally, we tested the last hypothesis that states that including atlas information jointly with the input image data can further improve segmentation accuracy and make the proposed method more robust. The test results for various configurations of the entire thalamus and each of the 13 nuclei, along with their average, are presented in Table 5. In this case, our initial assumption was partially validated, revealing that some nuclei were segmented more accurately using the atlas-based approach (smaller ones), while others did not benefit from the atlas approach. Since both methods exhibited no overall statistically significant differences but proved to be complementary, we decided not to favor one over the other. Instead, we adopted an ensemble approach by averaging their predictions. The resulting ensemble was clearly better than any of them yielding an average Dice coefficient of 0.9335, the nuclei with the highest dice being the Pulvinar Nucleus with a value of 0.9728 and the one with the lowest the Thalamus Mammillothalamic Tract, with a value of 0.8775 (the smallest nucleus).

*Table 5. Dice results for the different thalamic nuclei for the atlas and non-atlas versions as well as the ensemble of both. Mean volumes of each nucleus (in voxels) are also presented to relate them with their associated dice coefficient. *Represents statistically significant differences between No atlas and Atlas. $^{\alpha}$represents statistically significant differences between No atlas and Ensemble and $^{\beta}$ represents statistically significant differences between atlas and ensemble.*

| | No atlas | Atlas | Ensemble | Voxels |
|---|---|---|---|---|
| Whole thalamus | 0.9795 ± 0.0041 | 0.9803 ± 0.0037 | **0.9820 ± 0.0033**$^{\alpha\beta}$ | 32892 |
| *Anterior Ventral Nucleus* | 0.9169 ± 0.0298 | 0.9173 ± 0.0334 | **0.9221 ± 0.0307** $^{\alpha\ \beta}$ | 574 |
| *Ventral Anterior Nucleus* | 0.9211 ± 0.0220 | 0.9204 ± 0.0242 | **0.9262 ± 0.0225** $^{\alpha\ \beta}$ | 1440 |
| *Ventral Lateral Anterior Nucleus* | 0.9144 ± 0.2080 | 0.9179 ± 0.0255 | **0.9214 ± 0.0229** $^{\alpha}$ | 493 |
| Ventral Lateral Posterior Nucleus | 0.9578 ± 0.0090* | 0.9543 ± 0.0113 | **0.9594 ± 0.0101** $^{\alpha\ \beta}$ | 4518 |
| Ventral Posterior Lateral Nucleus | 0.9403 ± 0.0151* | 0.9299 ± 0.0190 | **0.9414 ± 0.0152** $^{\beta}$ | 1918 |
| *Pulvinar Nucleus* | 0.9721 ± 0.0052* | 0.9680 ± 0.0060 | **0.9728 ± 0.0047** $^{\beta}$ | 7294 |
| *Lateral Geniculate Nucleus* | 0.9081 ± 0.0258 | 0.9113 ± 0.0270 | **0.9160 ± 0.0250** $^{\alpha\ \beta}$ | 463 |
| *Medial Geniculate Nucleus* | 0.9385 ± 0.0145 | 0.9414 ± 0.0192 | **0.9447 ± 0.0163** $^{\alpha}$ | 427 |
| *Centromedian Nucleus* | 0.9505 ± 0.0134 | 0.9509 ± 0.0155 | **0.9533 ± 0.0135** $^{\alpha}$ | 704 |
| *Mediodorsal Nucleus* | 0.9719 ± 0.0073* | 0.9693 ± 0.0070 | **0.9730 ± 0.0064** $^{\beta}$ | 3188 |
| *Habenular Nucleus* | 0.9082 ± 0.0374 | 0.9123 ± 0.0270 | **0.9168 ± 0.0318** $^{\alpha}$ | 124 |
| *Mammillothalamic Tract* | 0.8705 ± 0.1076 | 0.8714 ± 0.0930 | **0.8775 ± 0.1000** $^{\beta}$ | 112 |
| *Intermediate Space* | 0.9028 ± 0.0138 | 0.9004 ± 0.0140 | **0.9107 ± 0.0122** $^{\alpha\ \beta}$ | 11636 |
| Average | 0.9287 ± 0.0456* | 0.9281 ± 0.0423 | **0.9335 ± 0.0425** $^{\alpha\ \beta}$ | |

## 3.5. Semi-supervised learning results

Finally, in our last experiment, we employed semi-supervised training to not only enhance the accuracy of the method, but also, and more crucially, to improve its robustness. The 4856 lifespan cases were incrementally segmented and added to the training dataset as previously described. The results for the HCP test set (mean of the 13 nuclei and for the whole thalamus) are shown in Figure 3. It can be observed that both whole thalamus and the average dice of the nuclei improved over time. Detailed results are shown in Table 6.

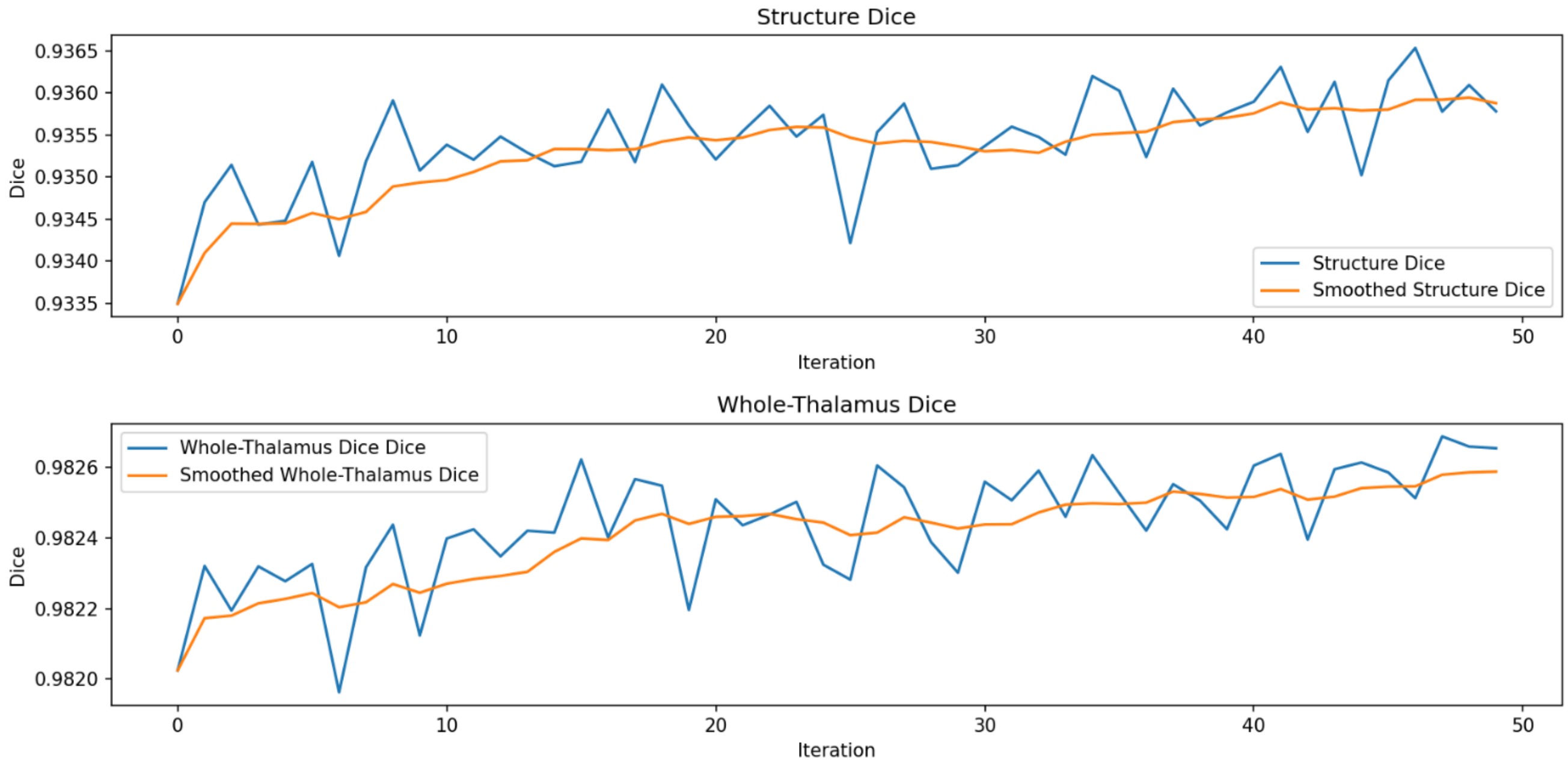

*Figure 3. Mean structure dice and whole thalamus dice at each iteration of the semi-supervised learning approach. The blue line indicates the average value and yellow its smoothed version.*

*Table 6. Comparison of the average Dice index of the ensemble method and the ensemble method trained with semi-supervised learning (applied to the DT test images) and the volume difference (VD). The best results are marked in bold. *Represents statistically significant differences.*

| | **Dice** | | **VD (%)** | |
|---|---|---|---|---|
| | **Ensemble** | **Semi-supervised model** | **Ensemble** | **Semi-supervised model** |
| Whole thalamus | 0.9820 ± 0.0033 | **0.9827 ± 0.0028*** | 0.90 ± 0.77 | 0.90 ± 0.73 |
| *Anterior Ventral Nucleus* | 0.9221 ± 0.0307 | **0.9268 ± 0.0279** | 5.45 ± 7.01 | **5.33 ± 6.43** |
| *Ventral Anterior Nucleus* | 0.9262 ± 0.0225 | **0.9342 ± 0.0215** | 3.00 ± 2.80 | **2.75 ± 2.51** |
| *Ventral Lateral Anterior Nucleus* | 0.9214 ± 0.0229 | **0.9272 ± 0.0210** | 6.07 ± 4.81 | **5.97 ± 4.92** |
| Ventral Lateral Posterior *Nucleus* | 0.9594 ± 0.0101 | **0.9618 ± 0.0087** | 1.70 ± 1.95 | **1.64 ± 1.94** |
| Ventral Posterior Lateral *Nucleus* | 0.9414 ± 0.0152 | **0.9421 ± 0.0153** | 2.29 ± 1.38 | **2.03 ± 1.52** |
| *Pulvinar Nucleus* | 0.9728 ± 0.0047 | **0.9735 ± 0.0050** | 1.40 ± 0.88 | **1.00 ± 0.70** |
| *Lateral Geniculate Nucleus* | 0.9160 ± 0.0250 | **0.9163 ± 0.0272** | **4.71 ± 3.95** | 5.33 ± 4.57 |
| *Medial Geniculate Nucleus* | 0.9447 ± 0.0163 | **0.9469 ± 0.0123** | 4.12 ± 3.01 | **3.83 ± 2.95** |
| *Centromedian Nucleus* | **0.9533 ± 0.0135** | 0.9517 ± 0.0122 | 3.44 ± 2.57 | **3.35 ± 2.84** |
| *Mediodorsal Nucleus* | 0.9730 ± 0.0064 | **0.9742 ± 0.0071** | **1.80 ± 1.56** | 2.09 ± 1.46 |
| *Habenular Nucleus* | 0.9168 ± 0.0318 | **0.9189 ± 0.0290** | **5.13 ± 3.49** | 5.15 ± 4.51 |
| *Mammillothalamic Tract* | 0.8775 ± 0.1000 | **0.8776 ± 0.1040** | **9.06 ± 11.13** | 10.06 ± 15.45 |
| *Intermediate Space* | 0.9107 ± 0.0122 | **0.9140 ± 0.0113*** | **2.64 ± 2.26** | 2.85 ± 2.07 |
| Average | 0.9335 ± 0.0425 | **0.9358 ± 0.0426*** | **3.91 ± 4.94** | 3.95 ± 5.90 |

As shown in Table 6, the dice index obtained is higher in the semi-supervised learning model except for one structure, the Centromedian Nucleus. This demonstrates the effectiveness of the semi-supervised approach applied to the test dataset and the usefulness of the approach of incorporating in order the most similar data until reaching the most difficult ones in the last iteration. No statistical differences were found for VD metric.

However, although the proposed semi-supervised training strategy has proved to improve test accuracy we don't know if it effectively improves model generalization on other data. To assess this point, we used the lifespan dataset. Since there is no ground truth in this dataset we used an auxiliary metric. Specifically, we used the normal controls of the lifespan dataset (N=2887) to build lifespan models for each structure. Our assumption was that lifespan models constructed after the semi-supervised learning would have lower dispersion (bounds size) than lifespan models using the original ensemble method before fine tuning (i.e. segmentation errors will increase volume estimation variability for each structure). To build the models we used the method described in [26].

For 19 of the 26 labels (left and right thalamus), see Figure 4, the volume bounds for the new model were lower than the bounds for the original segmentations (prior to the semi-supervised training). To statistically compare all the structures, we used the Wilcoxon test for related samples, to determine whether there are significant differences in the mean standard deviation between the two models. A correction was applied to control the false discovery rate (FDR) using the Benjamini-Hochberg method. The results were statistically significant for all 19 nuclei, demonstrating that our semi-supervised training reduces model volume variability which suggests that the proposed method is more reliable/robust. The final proposed method is, therefore, the one that includes semi-supervised learning.

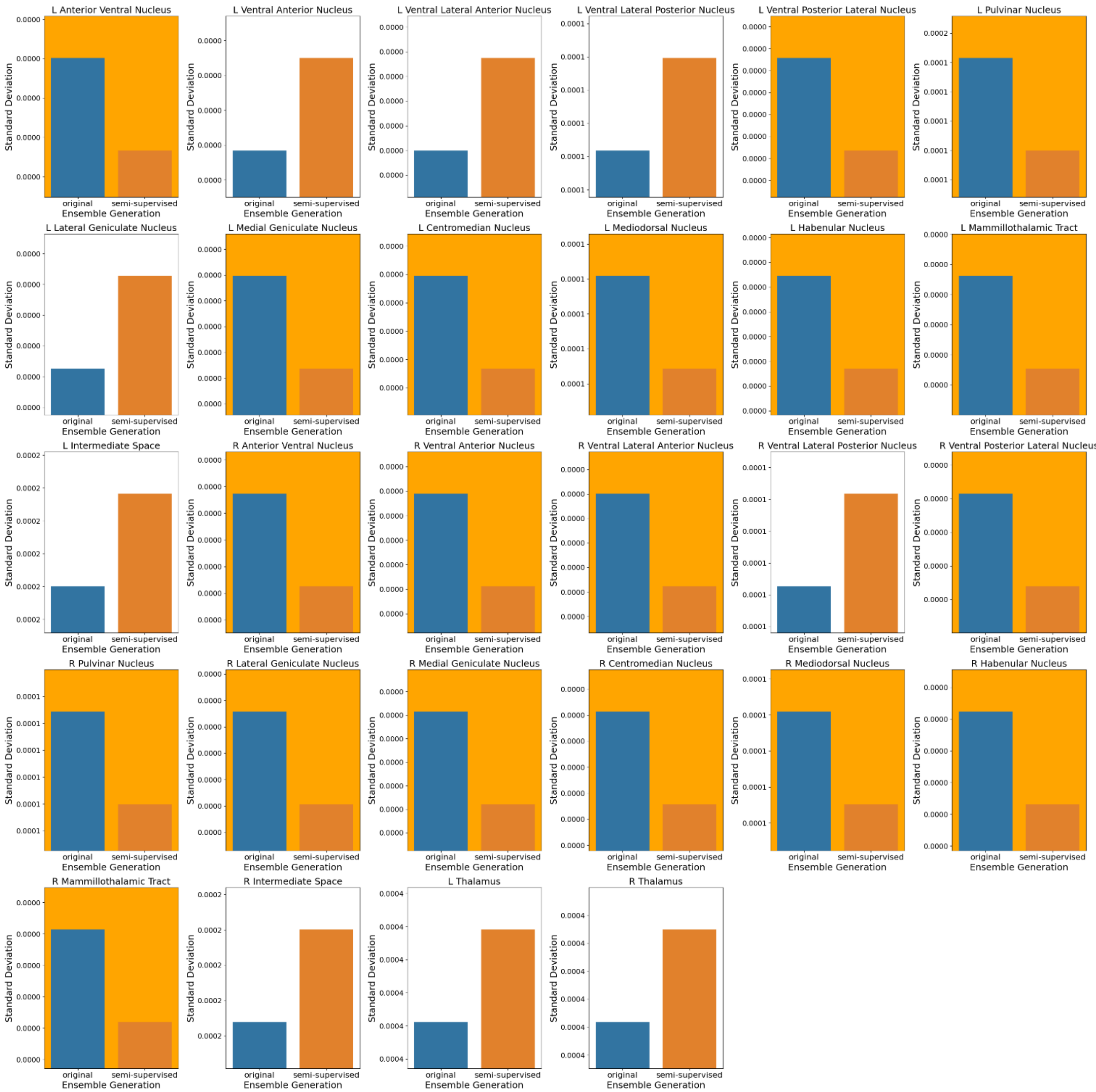

*Figure 4. Dispersion results of the two models, original (blue) and semi-supervised (orange) for each structure. Cases with statistically significant differences are marked with *. In orange background those where the new model reduces the dispersion (19/26 nuclei). Note that thalamic nuclei are not strictly symmetrical in terms of volume, partly due to the lateralization of the functions they process, such as auditory and visual inputs [43].*

## 3.6. State-of-the-art comparison

We perform a comparison of our final proposed method (including semi-supervised learning) with THOMAS method [21] as our labelling protocol is based on THOMAS protocol. Note that thalamic intermediate space was not included in the comparison as it is not defined in THOMAS protocol and the Mammillothalamic Tract has a different definition in our protocol. We also included the HIPS-THOMAS method [22] in the comparison which is a HIPS related method that allows to use T1 images as input.

We used the HCP test dataset, which consists of 20 T1 and synthesized WMn images. HIPS-THOMAS was applied to the 20 T1 images (please note that HIPS-THOMAS synthesizes internally the WMn images). THOMAS was used to the 20 synthesized WMn images via the available Docker container (https://github.com/thalamicseg/hipsthomasdocker). The results in Table 7 show that the THOMAS method outperforms HIPS-THOMAS in the Ventral Lateral Posterior Nucleus, Medial Geniculate Nucleus, Centromedian Nucleus, Mediodorsal Nucleus, and Habenular Nucleus. The findings, including Dice scores below 0.7 for the Habenular Nucleus and Ventral Lateral Anterior Nucleus, align with the observations in the referenced article [22]. The proposed method achieves higher Dice values across all structures. Table 8 presents the volume difference percentage, which is lower for HIPS-THOMAS compared to THOMAS, again in good agreement with the HIPS-THOMAS paper. Note that the volume difference percentage of the proposed method is significantly smaller than compared competing methods.

*Table 7. Average Dice results for the different thalamic nuclei for THOMAS, HIPS-THOMAS and the proposed method. Note that thalamic intermediate space and Mammillothalamic Tract are not not included in the comparison. The best results are in bold. *Represents statistically significant differences between Thomas and the proposed method. $^{\alpha}$represents statistically significant differences between HIPS-THOMAS and the proposed method.*

| **Thalamic nuclei** | **THOMAS** | **HIPS-THOMAS** | **Proposed** |
|---|---|---|---|
| Anterior Ventral Nucleus | 0.7632 ± 0.0685 | 0.7643 ± 0.0675 | **0.9268 ± 0.0279***$^{\alpha}$ |
| Ventral Anterior Nucleus | 0.7233 ± 0.0649 | 0.7253 ± 0.0642 | **0.9342 ± 0.0215***$^{\alpha}$ |
| Ventral Lateral Anterior Nucleus | 0.5587 ± 0.1069 | 0.6759 ± 0.0791 | **0.9272 ± 0.0210***$^{\alpha}$ |
| Ventral Lateral Posterior Nucleus | 0.8369 ± 0.0247 | 0.8273 ± 0.0286 | **0.9618 ± 0.0087***$^{\alpha}$ |
| Ventral Posterior Lateral Nucleus | 0.7399 ± 0.0566 | 0.7434 ± 0.0661 | **0.9421 ± 0.0153***$^{\alpha}$ |
| Pulvinar Nucleus | 0.8851 ± 0.0214 | 0.8896 ± 0.0166 | **0.9735 ± 0.0050***$^{\alpha}$ |
| Lateral Geniculate Nucleus | 0.7448 ± 0.0573 | 0.7467 ± 0.0347 | **0.9163 ± 0.0272***$^{\alpha}$ |
| Medial Geniculate Nucleus | 0.8421 ± 0.0559 | 0.7864 ± 0.0546 | **0.9469 ± 0.0123***$^{\alpha}$ |
| Centromedian Nucleus | 0.8021 ± 0.0331 | 0.7461 ± 0.0417 | **0.9517 ± 0.0122***$^{\alpha}$ |
| Mediodorsal Nucleus | 0.8674 ± 0.0519 | 0.8547 ± 0.0484 | **0.9742 ± 0.0071***$^{\alpha}$ |
| Habenular Nucleus | 0.7008 ± 0.1083 | 0.5972 ± 0.0947 | **0.9189 ± 0.0290***$^{\alpha}$ |
| **Average** | 0.7695 ± 0.1100 | 0.7597 ± 0.0973 | **0.9431 ± 0.0170***$^{\alpha}$ |

*Table 8. Volume difference (VD) percentage results for the different thalamic nuclei for THOMAS. HIPS-THOMAS and the proposed method. Note that that the thalamic intermediate space is not included in the comparison as it is not defined in THOMAS method and the* Mammillothalamic Tract has different definitions in the methods*. Best results in bold. *Represents statistically significant differences between Thomas and proposed method. $^{\alpha}$Represents statistically significant differences between HIPS-THOMAS and proposed method.*

| **Thalamic nuclei** | **THOMAS** | **HIPS-THOMAS** | **Proposed** |
|---|---|---|---|
| Anterior Ventral Nucleus | 37.51 ± 28.15 | 39.00 ± 25.06 | **5.33 ± 6.43***$^{\alpha}$ |
| Ventral Anterior Nucleus | 16.57 ± 19.41 | 13.26 ± 20.82 | **2.75 ± 2.51***$^{\alpha}$ |
| Ventral Lateral Anterior Nucleus | 18.72 ± 11.00 | 12.73 ± 9.54 | **5.97 ± 4.92***$^{\alpha}$ |
| Ventral Lateral Posterior Nucleus | 8.61 ± 5.39 | 10.94 ± 8.17 | **1.64 ± 1.94***$^{\alpha}$ |
| Ventral Posterior Lateral Nucleus | 17.39 ± 9.68 | 17.82 ± 10.71 | **2.03 ± 1.52***$^{\alpha}$ |
| Pulvinar Nucleus | 5.39 ± 4.46 | 4.55 ± 3.98 | **1.00 ± 0.70***$^{\alpha}$ |
| Lateral Geniculate Nucleus | 35.94 ± 22.74 | 26.75 ± 16.23 | **5.33 ± 4.57***$^{\alpha}$ |
| Medial Geniculate Nucleus | 16.81 ± 11.59 | 15.89 ± 9.13 | **3.83 ± 2.95***$^{\alpha}$ |
| Centromedian Nucleus | 9.91 ± 6.017 | 13.25 ± 8.12 | **3.35 ± 2.84***$^{\alpha}$ |
| Mediodorsal Nucleus | 14.47 ± 10.44 | 18.22 ± 10.32 | **2.09 ± 1.46***$^{\alpha}$ |
| Habenular Nucleus | 14.23 ± 17.70 | 22.50 ± 13.68 | **5.15 ± 4.51***$^{\alpha}$ |
| **Average** | 17.78 ± 18.06 | 17.72 ± 16.24 | **3.49 ± 3.12***$^{\alpha}$ |

## 3.7. DeepThalamus pipeline

As shown in this paper, the proposed method can accurately segment multimodal high-resolution T1/WMn images. However, we are aware that most MR images produced in research and clinical settings do not have the necessary resolution and/or multimodality. To make our method more accessible to the scientific community, we developed a complete pipeline that automatically performs the entire segmentation process using only standard-resolution T1 images, which are more commonly available. This pipeline, called DeepThalamus, is accessible through our online platform, volBrain (https://volbrain.net/services/DeepThalamus).

The proposed pipeline is based on an extensive preprocessing process aimed at preparing the data to be segmented. It consists of the following steps:

- **Noise removal**: The Spatially Adaptive Non-local means (SANLM) filter [27] was used to reduce random noise naturally present in the images.
- **Registration**: Affine registration to MNI152 space (1 mm isotropic resolution). ANTs software [29] was employed.
- **Inhomogeneity correction**: The N4 bias correction method was used to correct the inhomogeneity of the images [28].
- **Intensity normalization**: We normalized the T1 images applying a piecewise linear tissue mapping based on the TMS method [44] as described in the study by [9].
- **Intracranial cavity volume (ICV) extraction**: To compute normalized volumes, we segmented the ICV using the Deep ICE method [45].
- **Second inhomogeneity correction and intensity normalization**: This was performed using the ICV extracted volume instead of the original image to further improve the preprocessing.
- **Super-resolution**: The T1 image was super-resolved to 0.5 mm isotropic resolution (factor 2) using an in-house 3D ResNet-based super-resolution network trained using the full HCP dataset (N=1200). This step generated a T1 volume of 362x434x362 voxels.
- **Area of interest cropping**: Once the images were placed in MNI152 space, they were cropped to select only the sub volume containing the right and left thalamus. Cropped volumes were obtained using predefined limits based on the thalamus labels with a security margin of 10 voxels in each dimension to account for anatomical position variability. The final crop had a size of 76x91x79 voxels.
- **WMn synthesis**: The cropped T1 volumes were used to synthesize their WMn counterparts using a monomodal synthesis deep network like the one previously described but trained solely on cropped images.
- **Atlas creation**: Finally, a subject-specific atlas was created for each thalamus using the previously described method.

Once the cropped T1, synthetic WMn and atlas were obtained, the trained neural networks (with and without atlas) were applied and the ensemble of the results was used to generate the thalamic segmentation (using the described DPN architecture) for both left and right thalamus, each with their 13 labels.

Finally, a pdf report is generated with volumetric information of each thalamic nucleus, with information of its left/right asymmetry and normalized values related to the intracranial cavity volume. The prototype also includes normative bounds for each structure by sex and age (obtained from the lifespan dataset) so the results of each subject can be compared to their corresponding healthy lifespan model. The whole pipeline is summarized in Figure 5. The whole processing time of the pipeline is estimated to be around 3 minutes, making it very competitive and suitable for big data analysis.

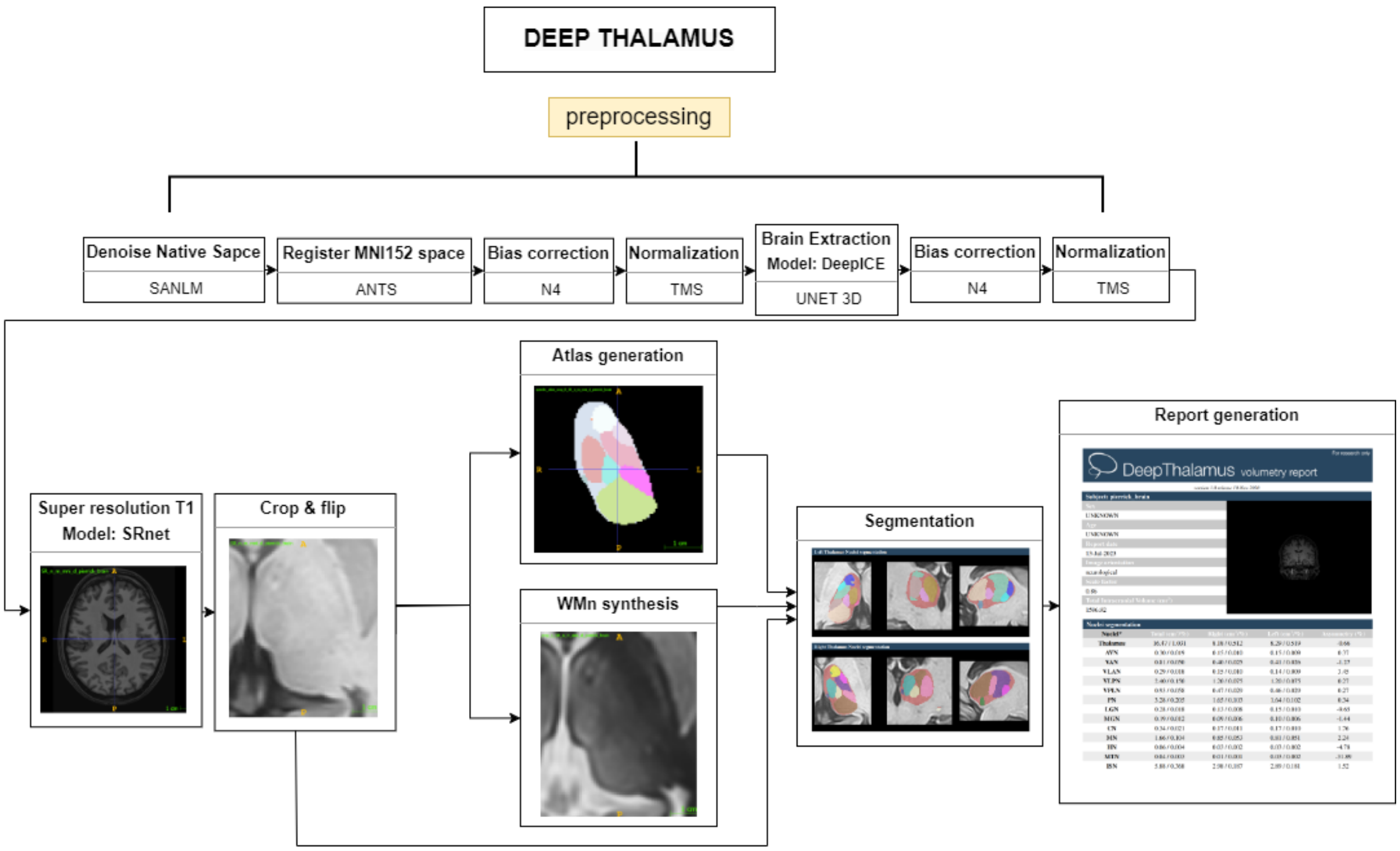

Figure 5. Scheme of the proposed DeepThalamus pipeline.

We are aware that results on super-resolved and synthetic WMn images are not going to be as good as in native high-quality data. Therefore, to estimate the performance drop, we used the proposed pipeline on monomodal and down-sampled T1 data from the HCP test set to estimate this drop. The results can be checked in Table 9. As can be noticed, there is a drop in the performance for the thalamic nuclei but not for the whole thalamus. However, even with the drop the results are still competitive in our opinion.

*Table 9. Dice results and volume difference (VD) for the monomodal simulated low quality T1 and multimodal HR data. The best results are marked in bold. *Represent statistically significant differences.*

| | **Dice** | | **VD (%)** | |
|---|---|---|---|---|
| | Monomodal LR | Multimodal HR | Monomodal LR | Multimodal HR |
| Whole thalamus | **0.9834 ± 0.0023** | 0.9827 ± 0.0028* | **0.81 ± 0.60** | 0.90 ± 0.73* |
| *Anterior Ventral Nucleus* | 0.9045 ± 0.0377 | **0.9268 ± 0.0279*** | 8.57 ± 10.48 | **5.33 ± 6.43** |
| *Ventral Anterior Nucleus* | 0.9043 ± 0.0247 | **0.9342 ± 0.0215*** | 5.12 ± 3.58 | **2.75 ± 2.5*** |
| *Ventral Lateral Anterior Nucleus* | 0.8890 ± 0.0381 | **0.9272 ± 0.0210*** | **6.51 ± 6.80** | 5.97± 4.92 |
| Ventral Lateral Posterior Nucleus | 0.9353 ± 0.0178 | **0.9618 ± 0.0087*** | 3.25 ± 3.31 | **1.64 ± 1.94** |
| Ventral Posterior Lateral Nucleus | 0.9030 ± 0.0345 | **0.9421 ± 0.0153*** | 4.42 ± 3.05 | **2.03 ± 1.52*** |
| *Pulvinar Nucleus* | 0.9622 ± 0.0066 | **0.9735 ± 0.0050*** | 2.43 ± 1.50 | **1.00 ± 0.70*** |
| *Lateral Geniculate Nucleus* | 0.9070 ± 0.0301 | **0.9163 ± 0.0272*** | 5.64 ± 6.18 | **5.33 ± 4.57** |
| *Medial Geniculate Nucleus* | 0.9270 ± 0.0132 | **0.9469 ± 0.0123*** | 5.19 ± 3.55 | **3.83 ± 2.95** |
| *Centromedian Nucleus* | 0.9200 ± 0.0231 | **0.9517 ± 0.0122*** | 4.72 ± 3.58 | **3.35 ± 2.84** |
| *Mediodorsal Nucleus* | 0.9556 ± 0.0129 | **0.9742 ± 0.0071*** | 5.45 ± 3.64 | **2.09 ± 1.46*** |
| *Habenular Nucleus* | 0.8713 ± 0.0625 | **0.9189 ± 0.0290*** | 8.73 ± 8.65 | **5.15 ± 4.51** |
| *Mammillothalamic Tract* | 0.8075 ± 0.1197 | **0.8776 ± 0.1040*** | 18.19 ± 25.95 | **10.06 ± 15.45*** |
| *Intermediate Space* | 0.8879 ± 0.0137 | **0.9140 ± 0.0113*** | 4.31 ± 2.31 | **2.85 ± 2.07*** |
| Average | 0.9058 ± 0.0580 | **0.9358 ± 0.0426*** | 6.35 ± 9.67 | **3.95 ± 5.90*** |

# 4. Discussion

In this paper, we have proposed an ultra-high resolution multimodal thalamic nuclei segmentation method. A novel architecture named DPN was proposed which improves the results of the well-known U-Net architecture despite its smaller size. We have also demonstrated that the higher resolution of the images is beneficial to better segment the thalamic nuclei, and that multimodality also helps in the segmentation process. We corroborated that the WMn images are the best suited for thalamic segmentation as previously pointed out in previous works [28,29,30].

We explored the use of a priori information in the form of a subject-specific thalamic atlas which helps to improve segmentation in smaller structures and proved to be complementary with non-atlas approaches. The proposed ensemble-based method was found to be the best performing option as segmentation strategy. To further improve the accuracy and generalization capabilities of the proposed method we used an incremental semi-supervised training method that successfully improved the results making it more suitable for analyzing MR data from different ages and anatomies.

The proposed method was compared with related state-of-the-art methods (THOMAS [21] and HIPS-THOMAS [22]) showing an improved performance for all the structures considered (0.77 vs 0.93). In terms of volume difference percentage, THOMAS performed better than HIPS-THOMAS in 6 out of 11 nuclei. However, the mean volume difference percentage for HIPS-THOMAS (17.72) was similar to THOMAS (17.78). Our proposed method demonstrated a better performance than competing methods for all nuclei and in average (3.49 vs 17).

We are aware of the limitations of the proposed approach. For example, the semi-automated nature of the training labelled dataset may produce more regular labels compared to manual labels which tend to be noisier and thus may have affected in the higher dice scores obtained by the proposed method (even including our exhaustive manual correction). Besides, even with the described atlas-based and semi-supervised learning used approach, the reduced number and diversity (young adults) of our labelled training dataset may still limit the generality of the proposed method.

Finally, a pipeline named DeepThalamus has been proposed able to process standard resolution T1 MR images making the method able to analyze legacy data without the need to acquire new sequences (WMn) or increase the image resolution (which will result in longer acquisition times usually not possible at clinical settings). This is done thanks to the use of advanced super-resolution and synthesis methods that simplify and make more accessible the proposed method to a wider range of users and acquisition settings. Furthermore, the reduced computational cost of the proposed pipeline (around 3 minutes) makes it ideal to process large databases efficiently, compared to [23], which takes around 45 minutes per case. Although the proposed DeepThalamus pipeline has a significant accuracy drop on standard resolution monomodal T1 data in comparison with the use of native multimodal HR data we believe that improvements in super-resolution and image synthesis methods can reduce this gap improving the overall accuracy of the proposed pipeline in the future.

## *5. Conclusion*

In this work, a new method for segmentation of thalamic nuclei based on deep learning using high-resolution multimodal MR images has been presented. Through a set of experiments, we have validated the hypothesis that using high resolution and multimodal data is beneficial to improve the accuracy of the segmentation of thalamic nuclei.

A new pipeline named DeepThalamus has been proposed and is publicly accessible to the whole scientific community through our online service volBrain (https://volbrain.net). This pipeline can work with usual standard 1 mm isotropic resolution T1 images which makes it very interesting to process many currently available datasets. We plan to use DeepThalamus to analyze the normal and pathological patterns of the thalamus to shed new light on this central structure which is involved in numerous neurological diseases.

## *6. Acknowledgments*

This work has been developed thanks to the project PID2020-118608RB-I00 and to the project PID2023-152127OB-I00 of the Ministerio de Ciencia, Innovacion y Universidades de España. This work also benefited from the support of the project DeepvolBrain and HoliBrain of the French National Research Agency (ANR-18-CE45-0013 and ANR-23-CE45-0020-01). This study was achieved within the Laboratory of Excellence TRAIL ANR-10-LABX-57 for the BigDataBrain project. Moreover, we thank the Investments for the future Program IdEx Bordeaux (ANR-10- IDEX- 03- 02, HL-MRI Project), Cluster of excellence CPU and the CNRS.

This work is based on multiple samples. We wish to thank all investigators of these projects who collected these datasets and made them freely accessible. The C-MIND data used in the preparation of this article were obtained from the C-MIND Data Repository (accessed in February 2015) created by the C-MIND study of Normal Brain Development. This is a multisite, longitudinal study of typically developing children from ages newborn through young adulthood conducted by Cincinnati Children's Hospital Medical Center and UCLA A listing of the participating sites and a complete listing of the study investigators can be found at https://research. cchmc.org/c-mind. The NDAR data used in the preparation of this manuscript were obtained from the NIH-supported National Database for Autism Research (NDAR). NDAR is a collaborative informatics system created by the National Institutes of Health to provide a national resource to support and accelerate research in autism. The NDAR dataset includes data from the NIH Pediatric MRI Data Repository created by the NIH MRI Study of Normal Brain Development. This is a multisite, longitudinal study of typically developing children from ages newborn through young adulthood conducted by the Brain Development Cooperative Group A listing of the participating sites and a complete listing of the study investigators can be found at http://pediatricmri.nih.gov/nihpd/info/participating_centers.html. Data collection and sharing for this project was funded by the Alzheimer's Disease Neuroimaging Initiative (ADNI) (National Institutes of Health Grant U01 AG024904) and DOD ADNI (Department of Defense award number W81XWH-12-2-0012). ADNI is funded by the National Institute on Aging, the National Institute of Biomedical Imaging and Bioengineering, and through generous contributions from the following: AbbVie, Alzheimer's Association; Alzheimer's Drug Discovery Foundation; Araclon Biotech; BioClinica, Inc.; Biogen; Bristol-Myers Squibb Company; CereSpir, Inc.; Cogstate; Eisai Inc.; Elan Pharmaceuticals, Inc.; Eli Lilly and Company; EuroImmun; F. Hoffmann-La Roche Ltd and its affiliated company Genentech, Inc.; Fujirebio; GE Healthcare; IXICO Ltd.; Janssen Alzheimer Immunotherapy Research & Development, LLC.; Johnson & Johnson Pharmaceutical Research & Development LLC.; Lumosity; Lundbeck; Merck & Co., Inc.; Meso Scale Diagnostics, LLC.; NeuroRx Research; Neurotrack Technologies; Novartis Pharmaceuticals Corporation; Pfizer Inc.; Piramal Imaging; Servier; Takeda Pharmaceutical Company; and Transition Therapeutics. The Canadian Institutes of Health Research is providing funds to support ADNI clinical sites in Canada. Private sector contributions are facilitated by the Foundation for the National Institutes of Health (www.fnih.org). The grantee organization is the Northern California Institute for Research and Education, and the study is coordinated by the Alzheimer's Therapeutic Research Institute at the University of Southern California. ADNI data are disseminated by the Laboratory for Neuro Imaging at the University of Southern California.

The OASIS data used in the preparation of this manuscript were obtained from the OASIS project. See http:// www.oasis-brains.org/for more details. The AIBL data used in the preparation of this manuscript were obtained from the AIBL study of ageing. See www.aibl.csiro.au for further details. The ICBM data was used in the preparation of this manuscript. The IXI data used in the preparation of this manuscript were supported by the - http://www. brain-development.org/. The ABIDE data used in the preparation of this manuscript were supported by ABIDE funding resources listed at http://fcon_1000.projects.nitrc.org/indi/abide/. ABIDE primary support for the work by Adriana Di Martino. Primary support for the work by Michael P. Milham and the INDI team was provided by gifts from Joseph P. Healy and the Stavros Niarchos Foundation to the Child Mind Institute. http://fcon_1000.projects.nitrc.org/indi/abide/. This manuscript reflects the views of the authors and may not reflect the opinions or views of the database providers.

## 7. Ethics statement

This manuscript uses de-identified MRI scans from publicly available databases. All participants provided written informed consent according to the enrolment protocols of their respective open data sources. To maintain privacy and confidentiality, no personal identifying information was included in the analysis.